\documentclass[conference,letterpaper,10pt]{IEEEtran}

\usepackage[utf8]{inputenc}
\usepackage[T1]{fontenc}
\usepackage{amsmath,amssymb,amsfonts}
\usepackage{algorithmic}
\usepackage{algorithm}
\usepackage{graphicx}
\usepackage{textcomp}
\usepackage{xcolor}
\usepackage{booktabs}
\usepackage{multirow}
\usepackage{tabularx}
\usepackage{url}
\usepackage{hyperref}
\usepackage{cite}
\usepackage{listings}
\usepackage{enumitem}
\usepackage{subcaption}
\usepackage{balance}
\usepackage{tikz}
\usetikzlibrary{positioning,arrows.meta,fit,backgrounds,calc}

\hypersetup{
  colorlinks=true,
  linkcolor=blue!70!black,
  citecolor=blue!70!black,
  urlcolor=blue!70!black
}

\newcommand{\eg}{e.g.}

\newcommand{\etal}{et al.}

\title{AESP: A Human-Sovereign Economic Protocol for AI Agents with Privacy-Preserving Settlement}

\author{
  \texttt{Jian Sheng Wang} \\
  \texttt{jason@acechain.io} \\
  Montreal, Canada
}
\date{November 24, 2026}

\begin{document}

\maketitle

\begin{abstract}
As AI agents increasingly perform economic tasks on behalf of humans, agent autonomy comes into conflict with human control over financial assets. We present the \textbf{Agent Economic Sovereignty Protocol (AESP)}, a layered protocol in which agents transact autonomously at machine speed on crypto-native infrastructure while remaining cryptographically bound to human-defined governance boundaries. Under a deployment assumption that agents reach the signing module only through its policy engine, AESP enforces the invariant that \emph{agents are economically capable but never economically sovereign} through five mechanisms: (1)~a deterministic eight-check policy engine with tiered escalation; (2)~human-in-the-loop review with automatic, explicit, and biometric tiers; (3)~EIP-712 dual-signed commitments that an external escrow contract can verify; (4)~HKDF-based context-isolated privacy with batched consolidation; and (5)~an ACE-GF-based cryptographic substrate. We formalize two testable hypotheses on security coverage and latency overhead, and specify an evaluation methodology with baselines and an ablation design. The protocol is implemented as an open-source TypeScript SDK (208 tests, ten modules) with interoperability via MCP and A2A.\footnote{\url{https://github.com/ya-xyz/aesp}}
\end{abstract}


\section{Introduction}
\label{sec:intro}

\subsection{Motivation}

Large language model (LLM)-based agents can reason, plan, and execute multi-step tasks with minimal human guidance~\cite{ouyang2022instructgpt}. When these tasks involve financial transactions (purchasing goods, subscribing to services, settling escrow), two requirements come into conflict. On one hand, agents must be \emph{autonomous enough} to act without constant human micro-management; a human approving every API call defeats the purpose of delegation. On the other hand, humans must retain \emph{sovereignty} over when and how their money moves; an agent that can unilaterally drain a wallet is an unacceptable liability.

Existing wallet and DeFi interfaces are designed for direct human interaction. Agent-facing protocols that enforce human-defined economic boundaries are largely absent from the current ecosystem. Recent surveys of autonomous agents on blockchains~\cite{alqithami2026survey} identify trust boundaries as a high-value attack surface, and the emerging ``agent economy'' literature~\cite{xu2026agenteconomy} proposes architectures for agent-to-agent commerce but often assumes full agent autonomy rather than human-sovereign control. The tension between autonomy and accountability is a recognized open problem in decentralized AI systems~\cite{hu2025trustless, hu2026sovereign}.

The literature distinguishes two models of AI economic participation: \emph{fully autonomous} agents holding their own wallets on permissionless rails, and \emph{human-delegated} agents operating strictly as legally accountable extensions of a specific human principal. We argue for, and implement, a third option, which we call \emph{sovereign agent economics}: agents transact autonomously at machine speed on crypto-native infrastructure while remaining cryptographically bound to human-defined governance boundaries. The degree of autonomy then becomes a set of tunable policy parameters.

Most recently, Toma\v{s}ev~\etal{}~\cite{tomasev2026delegation} from Google DeepMind proposed a conceptual framework for ``Intelligent AI Delegation,'' identifying nine dimensions of safe delegation, including permission handling, verifiable task completion, trust and reputation, and monitoring. The framework provides a taxonomy and design vocabulary; the authors note that concrete implementations realizing these principles remain an open challenge.

\subsubsection*{Gap}
Existing frameworks~\cite{tomasev2026delegation, cheng2026threepillar, hu2026sovereign} identify the requirements for safe economic delegation but provide no implementation or empirical validation. No prior work quantifies the trade-off between sovereignty enforcement and transaction efficiency in agent economic systems. Specifically, no existing system answers: \emph{What latency, false-positive, and usability cost does enforcing human control over every agent economic action impose?}

\subsubsection*{Hypotheses}
AESP addresses this gap. We formalize our claims as two testable hypotheses:

\begin{itemize}[leftmargin=*]
\item \textbf{H1 (Security):} AESP's eight-check policy gate automatically blocks at least 90\% of unauthorized transactions while auto-blocking or escalating at most 5\% of legitimate requests. Escalation is evaluated separately as operational burden (review load), not as algorithmic security success.

\item \textbf{H2 (Efficiency):} The end-to-end latency overhead introduced by AESP (policy evaluation, cryptographic signing via the Rust module, and HKDF key derivation) remains below 200\,ms median per transaction and does not materially degrade transaction completion rates compared to an unconstrained baseline.
\end{itemize}

AESP is an implemented protocol layer between the human's \emph{Digital Sovereign Entity} (DSE), i.e., the human and their devices, and the settlement layer (vaults, escrow, allowances). Every agent action is either permitted by policy or escalated to the human for approval. The protocol draws on established access control models~\cite{sandhu1996rbac, hu2014abac, dennis1966capability}, cryptographic commitment schemes~\cite{eip712, rfc5869}, and blockchain privacy techniques~\cite{erc5564, tornado, wicht2024privacy}. Section~\ref{sec:evaluation} specifies the evaluation methodology.

\subsection{Design Principle}

The central invariant of AESP is:

\begin{quote}
\textbf{Agents should be economically capable but never economically sovereign.}
\end{quote}

\noindent This invariant is enforced through five mechanisms:

\begin{enumerate}[leftmargin=*]
\item \textbf{Policy-gated execution.} Every agent action is evaluated against a deterministic sequence of eight policy checks (per-transaction limit, time window, address allowlist, chain allowlist, method allowlist, first-payment review, minimum balance, and budget limits) before execution is permitted (Section~\ref{sec:policy}).

\item \textbf{Human-in-the-loop review.} Actions that fail any policy check are routed to the human's device for explicit approval. Critical policy changes, such as increasing spending limits or broadening agent scope, require biometric confirmation, following the progressive validation model advocated by recent AI safety frameworks~\cite{cheng2026threepillar} (Section~\ref{sec:review}).

\item \textbf{Cryptographic commitment.} Agent-to-agent agreements are structured as EIP-712~\cite{eip712} typed data, signed by both buyer and seller agents, and optionally backed by on-chain escrow. The dual-signing requirement ensures that neither party can unilaterally modify the terms after commitment (Section~\ref{sec:commitment}).

\item \textbf{Context-isolated privacy.} Each transaction uses an HKDF-derived~\cite{rfc5869} ephemeral address whose derivation context includes the agent ID, direction, sequence number, and transaction ID. Different contexts produce cryptographically independent addresses, so the addresses themselves do not let on-chain observers correlate transactions across agent contexts; consolidating funds back to the vault re-links them (Section~\ref{sec:privacy}).

\item \textbf{Cryptographic execution substrate.} The ACE-GF module~\cite{acegf2026} provides multi-chain signing, Argon2id/HKDF-based derivation, context-isolated address generation, and optional ML-DSA-44 support across WASM and native FFI targets. Under deployment assumption DA1, key material remains within the cryptographic module boundary (Section~\ref{sec:crypto}).
\end{enumerate}

\subsection{Contributions}

This paper makes the following contributions:

\begin{itemize}[leftmargin=*]
\item A \textbf{protocol specification} with formal definitions of the agent identity model, policy evaluation algorithm, negotiation finite state machine, commitment lifecycle, review escalation flow, and privacy derivation scheme (Sections~\ref{sec:identity}--\ref{sec:privacy}).

\item A \textbf{cryptographic layer} based on the ACE-GF substrate~\cite{acegf2026} (a Rust implementation with WASM and native FFI bindings), providing multi-curve signing (Ed25519, secp256k1, ML-DSA-44), authenticated encryption, Argon2id-to-HKDF key derivation, and context-isolated multi-chain address generation via the REV32 wallet format (Section~\ref{sec:crypto}).

\item An \textbf{open-source SDK implementation} in TypeScript with 208 tests across ten modules,\footnote{\url{https://github.com/ya-xyz/aesp}} covering identity derivation, policy enforcement, negotiation, commitments, review, MCP tools, A2A cards, cryptographic primitives, and context-isolated privacy (Section~\ref{sec:implementation}).

\item An \textbf{evaluation framework} comprising four baselines of increasing restrictiveness, a per-check ablation design, an unlinkability security game, and two falsifiable hypotheses on security coverage and latency overhead (Section~\ref{sec:evaluation}).

\item A \textbf{privacy design} with HKDF-derived ephemeral addresses, batched consolidation with timing jitter and Fisher-Yates shuffle, targeting transaction unlinkability against practical on-chain analysis while preserving auditability (Section~\ref{sec:privacy}).

\item \textbf{Interoperability} with existing AI frameworks via the Model Context Protocol~\cite{mcp2025} (eight tool definitions) and Google's Agent-to-Agent protocol~\cite{a2a2025} (agent card generation) (Section~\ref{sec:interop}).
\end{itemize}

\subsection{Relationship to Existing Frameworks}

We position this work relative to the ``Intelligent AI Delegation'' framework by Toma\v{s}ev~\etal{}~\cite{tomasev2026delegation}. Table~\ref{tab:deepmind_comparison} provides a detailed mapping between the conceptual requirements identified in that framework and the corresponding mechanisms implemented in AESP. AESP's design draws on their taxonomy of delegation dimensions (permission handling, monitoring, verifiable task completion, trust and reputation) and instantiates it as a policy engine with 8 deterministic checks, a human-in-the-loop review queue, EIP-712 dual-signed commitments, and cryptographic identity certificates. The two works are complementary.

\begin{table*}[t]
\centering
\caption{Mapping between Google DeepMind's Intelligent AI Delegation Framework~\cite{tomasev2026delegation} and AESP Implementation}
\label{tab:deepmind_comparison}
\small
\begin{tabularx}{\textwidth}{@{}lXX@{}}
\toprule
\textbf{Delegation Dimension} & \textbf{Conceptual Requirement~\cite{tomasev2026delegation}} & \textbf{AESP Realization} \\
\midrule
Permission Handling (\S4.7) & Tiered permission models with progressive trust & PolicyEngine: 8-check evaluation, 3 approval tiers (auto/review/biometric), 8 critical change types \\
\addlinespace
Monitoring (\S4.5) & Continuous monitoring and logging & ReviewManager: async review queue with priority, expiration deadlines, 6 event types, emergency freeze \\
\addlinespace
Verifiable Completion (\S4.8) & Verifiable task completion mechanisms & CommitmentBuilder: EIP-712 dual-signed commitments, 7-state lifecycle, SHA-256 commitment hashes \\
\addlinespace
Trust \& Reputation (\S4.6) & Trust establishment and reputation models & Identity certificates with owner Ed25519 signatures, hierarchical delegation (depth $\leq 5$), DID-based identity \\
\addlinespace
Security (\S4.9) & Cryptographic security and key management & ACE-GF substrate (Section~\ref{sec:crypto}; WASM/C FFI/Dart FFI): Ed25519, secp256k1, X25519, ML-DSA-44, AES-256-GCM; keys never leave crypto boundary \\
\addlinespace
Agent Communication & Standardized agent communication protocols & 8 MCP tools + A2A agent cards (implemented, tested) \\
\addlinespace
Privacy & Not explicitly addressed & HKDF context-isolated ephemeral addresses, Fisher-Yates shuffle, ${\pm}$30\% timing jitter, batched consolidation \\
\addlinespace
Post-Quantum Readiness & Not addressed & ML-DSA-44 (FIPS 204) lattice-based signatures via cryptographic module \\
\bottomrule
\end{tabularx}
\end{table*}

\subsection{Outline}

Section~\ref{sec:related} surveys related work. Section~\ref{sec:model} defines the system model, architecture, and threat model. Sections~\ref{sec:identity}--\ref{sec:privacy} present the protocol components: identity, cryptographic foundation, policy, negotiation, commitment, review, and privacy. Section~\ref{sec:implementation} describes the implementation. Section~\ref{sec:interop} discusses interoperability. Section~\ref{sec:evaluation} presents the evaluation methodology and experimental design for validating H1 and H2. Section~\ref{sec:casestudies} illustrates protocol expressiveness through case studies. Section~\ref{sec:discussion} addresses limitations. Section~\ref{sec:conclusion} concludes.

\section{Related Work}
\label{sec:related}

\subsection{Agent Economic Frameworks}

The idea of autonomous economic agents predates LLMs. Minarsch~\etal{}~\cite{minarsch2020aea} introduced the Fetch.ai Autonomous Economic Agent (AEA) framework, where software agents pursue economic goals using a decentralized Open Economic Framework with blockchain settlement. More recently, Xu~\cite{xu2026agenteconomy} proposed a five-layer architecture for ``The Agent Economy'' spanning decentralized physical infrastructure (DePIN), decentralized identifiers (DIDs), cognitive tooling, ERC-4337 settlement, and Agentic DAOs. Their design assumes full agent autonomy: agents possess independent economic agency. AESP addresses the same problem space but takes the opposite stance: agents operate \emph{under} human sovereignty, not independently of it.

Alqithami~\cite{alqithami2026survey} surveys 317 works on autonomous agents on blockchains, proposing a five-part taxonomy of integration patterns and identifying trust boundaries as a primary attack surface. Hu~\etal{}~\cite{hu2025trustless} analyze the paradox of deploying LLM-based agents on trustless substrates (blockchain, TEEs), where immutability gains tamper-resistance but loses oversight mechanisms; AESP's policy-gated escalation model targets this tension.

\subsection{Intelligent AI Delegation}

Toma\v{s}ev~\etal{}~\cite{tomasev2026delegation} present a conceptual treatment of AI delegation. Their framework identifies eleven task characteristics, defines delegation as a principal-agent relationship, and proposes nine dimensions of safe delegation: task specification, resource provisioning, tool access, monitoring, trust and reputation, permission handling, verifiable task completion, security, and rollback. The taxonomy gives a vocabulary for reasoning about agent delegation; the authors identify implementation as an open direction. Table~\ref{tab:deepmind_comparison} maps each framework dimension to the corresponding AESP mechanism, illustrating how conceptual requirements can be realized as protocol components.

\subsection{Human-in-the-Loop AI Control}

Learning from human feedback was established by Christiano~\etal{}~\cite{christiano2017rlhf}, who showed that complex RL tasks can be solved using non-expert human preference feedback on fewer than 1\% of agent interactions. Ouyang~\etal{}~\cite{ouyang2022instructgpt} extended this to language models with InstructGPT, demonstrating that RLHF produces outputs aligned with human preferences.

At the systems level, Cheng~\etal{}~\cite{cheng2026threepillar} propose a three-pillar model (transparency, accountability, trustworthiness) with progressive validation analogous to staged autonomous driving. AESP's tiered escalation (routine transactions are auto-approved, unusual transactions require explicit review, and critical policy changes require biometric confirmation) applies this progressive approach to economic transactions. Hu and Rong~\cite{hu2026sovereign} identify the accountability gap that arises when cryptographic self-custody scaffolds agentic sovereignty; AESP addresses this by keeping the human principal as the ultimate economic authority.

\subsection{Policy and Authorization Models}

AESP's policy engine draws on three traditions in access control. Role-Based Access Control (RBAC)~\cite{sandhu1996rbac} assigns permissions to roles rather than individuals; AESP adapts this by assigning spending authority to agent \emph{scopes}. Attribute-Based Access Control (ABAC)~\cite{hu2014abac} evaluates policies over subject, object, operation, and environment attributes; AESP's context-aware policy checks follow this pattern. Capability-based security~\cite{dennis1966capability} grants unforgeable tokens for specific rights; AESP's per-policy, time-limited spending authorizations serve an analogous function.

Recent work has begun adapting these models for AI agents. Abaev~\etal{}~\cite{abaev2026agentguardian} present AgentGuardian, which monitors agent execution traces to learn legitimate behaviors and derive adaptive access-control policies. Ganie~\cite{ganie2025rbac} proposes integrating RBAC into LLM-based agents for industrial applications. AESP's contribution is a hybrid approach: static policies defined by the human principal, with tiered escalation for out-of-policy actions and a classification system for detecting critical policy changes.

\subsection{Blockchain Privacy}

Transaction privacy on public blockchains is well-studied. Stealth addresses (ERC-5564)~\cite{erc5564} allow recipients to receive assets without publicly linking transactions. Mixing protocols such as Tornado Cash~\cite{tornado} use zero-knowledge proofs, while CoinJoin~\cite{coinjoin} combines multiple inputs and outputs. Wicht~\etal{}~\cite{wicht2024privacy} formalize blockchain privacy notions (untraceability and unlinkability) using a Transaction DAG model.

AESP's privacy mechanism differs in that it does not require mixing or zero-knowledge proofs. Instead, it uses REV32 context-isolated derivation, where chain-specific key material is derived from an identity root via HKDF domain separation and per-transaction context labels~\cite{rfc5869}. The resulting addresses are cryptographically independent, providing address-level unlinkability without on-chain mixing; residual side-channel leakage from consolidation patterns is discussed in Section~\ref{sec:privacy}.

\subsection{Cryptographic Commitments and Escrow}

EIP-712~\cite{eip712} defines a standard for hashing and signing typed structured data with domain separators that prevent cross-application signature replay. AESP uses EIP-712 for its dual-signed payment commitments. Asgaonkar and Krishnamachari~\cite{asgaonkar2019escrow} propose dual-deposit escrow smart contracts for provably cheat-proof delivery; AESP's commitment model is compatible with such designs, though the protocol is settlement-layer agnostic.

\subsection{Agent Interoperability}

The Model Context Protocol (MCP)~\cite{mcp2025} provides a standardized interface for connecting AI assistants to external tools via JSON-RPC. Google's Agent-to-Agent (A2A) protocol~\cite{a2a2025} enables cross-framework agent discovery through Agent Cards. AESP exposes its economic operations as MCP tools and generates A2A agent cards, enabling any compatible framework to discover AESP-mediated services.

\subsection{Digital Sovereignty and Self-Sovereign Identity}

The concept of self-sovereign identity (SSI), articulated by Allen~\cite{allen2016ssi} and standardized through W3C DIDs~\cite{w3cdid} and Verifiable Credentials~\cite{w3cvc}, holds that individuals should control their own digital identities. Krul~\etal{}~\cite{krul2024ssi} systematize trust models across SSI components. Srivastava and Bullock~\cite{srivastava2024digital} examine digital sovereignty in the AI context. AESP applies the SSI principle of user control to agent spending.

\subsection{Post-Quantum Cryptography}

NIST's selection of ML-DSA (formerly CRYSTALS-Dilithium) as the primary post-quantum digital signature standard under FIPS~204~\cite{fips204} has motivated research into post-quantum readiness for blockchain systems. Fernandez-Carames and Fraga-Lamas~\cite{fernandez2020pqblockchain} survey post-quantum approaches for blockchain. AESP's cryptographic module includes ML-DSA-44 as an optional scheme for agent identity and transaction signing.

\section{System Model and Architecture}
\label{sec:model}

\subsection{Layered Architecture}

AESP is organized as a four-layer stack (Figure~\ref{fig:architecture}):

\begin{figure}[h]
\centering
\begin{tikzpicture}[
  layer/.style={draw, rounded corners=3pt, minimum width=0.88\columnwidth,
    minimum height=1.1cm, align=center, font=\small},
  lbl/.style={font=\scriptsize\sffamily, anchor=east},
  arr/.style={-{Stealth[length=5pt]}, thick, gray!70},
  arrdown/.style={-{Stealth[length=5pt]}, thick, blue!50!black},
  arrup/.style={-{Stealth[length=5pt]}, thick, red!50!black},
]
\node[layer, fill=blue!8] (L4)
  {\textbf{DSE (Digital Sovereign Entity)}\\[-1pt]
   {\scriptsize Human principal\;\textbullet\;Mobile\;\textbullet\;Browser extension}};
\node[layer, fill=green!6, below=0.35cm of L4] (L3)
  {\textbf{AESP Protocol}\\[-1pt]
   {\scriptsize Identity\;\textbar\;Policy\;\textbar\;Negotiation\;\textbar\;Commitment\;\textbar\;Review\;\textbar\;Privacy\;\textbar\;Crypto}};
\node[layer, fill=orange!8, below=0.35cm of L3] (L2)
  {\textbf{Interoperability Bridge}\\[-1pt]
   {\scriptsize MCP tools (8)\;\textbar\;A2A agent cards}};
\node[layer, fill=gray!8, below=0.35cm of L2] (L1)
  {\textbf{Settlement Layer}\\[-1pt]
   {\scriptsize Vaults\;\textbar\;Escrow\;\textbar\;Allowances\;\textbar\;Authority}};
\node[lbl] at ($(L4.west)+(-0.15,0)$) {\textbf{L4}};
\node[lbl] at ($(L3.west)+(-0.15,0)$) {\textbf{L3}};
\node[lbl] at ($(L2.west)+(-0.15,0)$) {\textbf{L2}};
\node[lbl] at ($(L1.west)+(-0.15,0)$) {\textbf{L1}};
\draw[arrdown] (L4.south) -- (L3.north)
  node[midway, right, font=\scriptsize, text=blue!50!black] {policy, freeze};
\draw[arrdown] (L3.south) -- (L2.north)
  node[midway, right, font=\scriptsize, text=blue!50!black] {signed auth};
\draw[arrdown] (L2.south) -- (L1.north)
  node[midway, right, font=\scriptsize, text=blue!50!black] {execute};
\draw[arrup] ($(L3.north)+(0.15,0)$) -- ($(L4.south)+(0.15,0)$)
  node[midway, left, font=\scriptsize, text=red!50!black] {review};
\draw[arrup] ($(L2.north)+(0.15,0)$) -- ($(L3.south)+(0.15,0)$)
  node[midway, left, font=\scriptsize, text=red!50!black] {status};
\draw[arrup] ($(L1.north)+(0.15,0)$) -- ($(L2.south)+(0.15,0)$)
  node[midway, left, font=\scriptsize, text=red!50!black] {events};
\end{tikzpicture}
\caption{AESP four-layer architecture. Blue arrows (downward) represent the sovereignty and authorization flow; red arrows (upward) represent escalation and event reporting.}
\label{fig:architecture}
\end{figure}
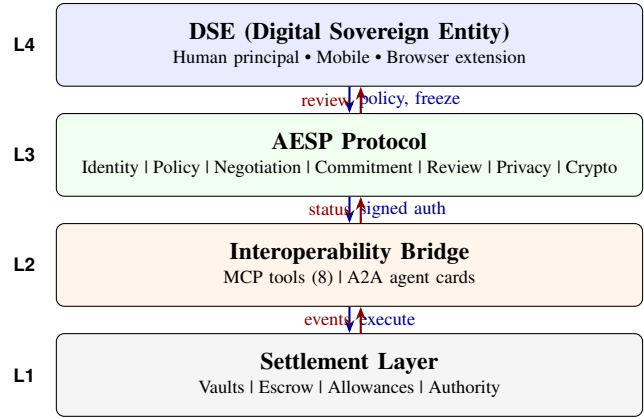

\textbf{Layer~4 (DSE)} represents the human principal and their devices (mobile phone, browser extension). The DSE holds the master ACE-GF mnemonic (REV32-encoded sealed entropy), defines policies, and approves or rejects escalated actions.

\textbf{Layer~3 (AESP)} implements the protocol logic: agent identity derivation and certification, policy evaluation, negotiation state management, commitment construction, review queue management, cryptographic operations (via an ACE-GF-based cryptographic substrate), and privacy-preserving address derivation. AESP does not hold funds or execute on-chain transactions; it produces signed authorizations and commitments.

\textbf{Layer~2 (Bridge)} exposes AESP operations to external AI frameworks. MCP tool definitions allow any MCP-compatible agent to check balances, create allowances, or file disputes. A2A agent cards advertise agent capabilities for cross-framework discovery.

\textbf{Layer~1 (Settlement)} performs on-chain execution: funding escrow, releasing payments, managing vaults and allowances. AESP is settlement-layer agnostic; any conforming implementation can serve as Layer~1.

\subsection{Participants}

We define three categories of participants:

\begin{itemize}[leftmargin=*]
\item \textbf{Human principal}~$H$: The owner of the master ACE-GF mnemonic and the ultimate economic authority. $H$ defines policies, approves escalated actions, and can freeze any agent at any time.
\item \textbf{Agent}~$A_i$: A software entity derived from $H$'s master key at index $i$, identified by $\mathrm{id}(A_i) = \mathrm{SHA\text{-}256}(\mathrm{pk}(A_i))$, operating within the policy boundaries set by $H$.
\item \textbf{Counterparty agent}~$A_j$: An agent controlled by a different human principal $H'$, with whom $A_i$ may negotiate and form commitments.
\end{itemize}

\subsection{Threat Model}
\label{sec:threat}

We formalize the threat model in terms of adversary capabilities, defense guarantees, and explicit failure modes.

\subsubsection{Adversary Capabilities}

\textbf{$\mathcal{A}_1$: Compromised agent.} The adversary gains control of agent $A_i$'s runtime (\eg, through prompt injection, supply-chain compromise, or software vulnerability). Formally, $\mathcal{A}_1$ can invoke any SDK function available to $A_i$ and can craft arbitrary action requests $r$ with any parameters (amount, recipient, chain, method).

\textbf{Deployment assumption (DA1: Mandatory policy mediation).} We assume that the deployment architecture enforces that \emph{all} agent signing requests are routed through the AESP SDK's PolicyEngine before reaching the cryptographic module. Concretely, the signing functions are not directly exposed to the agent runtime; they are accessible only through the SDK's policy-gated API. This is an architectural invariant that must be enforced by the host environment. The reference deployment pattern for browser environments loads the ACE-GF WASM module, the PolicyEngine, and the BudgetTracker in a dedicated Web Worker, so policy evaluation and signing share one isolation boundary. The Worker accepts only action requests over \texttt{postMessage}: it parses the payload to be signed, evaluates policy on the parsed fields (recipient, value, token, method, allowance) rather than on values declared by the caller, chooses the signing context itself, and signs only if the request is approved. It does not expose a raw \texttt{sign} operation, a caller-selected derivation context, or signing of an opaque hash; requests whose payload cannot be parsed are escalated to review. Raw credential material (mnemonic, passphrase) is passed once at initialization and never returned to the caller. The agent runtime on the main thread has no direct access to WASM linear memory or signing exports. In Node.js, the equivalent isolation is achieved via \texttt{worker\_threads} or a subprocess with IPC; in native FFI deployments, a sandboxed execution context provides the same boundary. If $\mathcal{A}_1$ can bypass the SDK and invoke signing functions directly, the policy enforcement guarantee is void. Under DA1, $\mathcal{A}_1$ cannot modify the cryptographic binary, access the master ACE-GF mnemonic material, or bypass the PolicyEngine evaluation path. DA1 also covers policy state. Policies, budget counters and reservations, first-payment records, and freeze status are held by the Worker and persisted only to storage the agent runtime cannot write (\eg, a separate origin or extension context in browsers, or a file owned by a separate OS account on servers); \texttt{localStorage} or IndexedDB in the agent's own origin does not satisfy DA1. The Worker accepts a policy only if its hash matches $h(\pi)$ in a certificate signed by $\mathrm{xid}_H$ (Section~\ref{sec:identity}), and it acts on a review decision, policy change, freeze, or unfreeze only if the message is signed by $\mathrm{xid}_H$ from the DSE. The agent can request these operations but cannot perform them.

\textbf{$\mathcal{A}_2$: Malicious counterparty.} Agent $A_j$ (or its principal $H'$) may manipulate negotiation outcomes, repudiate commitments, or exploit timing in the escrow flow. AESP mitigates this through the dual-signing commitment protocol (Section~\ref{sec:commitment}): both parties must sign the same EIP-712 structure before escrow is funded.

\textbf{$\mathcal{A}_3$: On-chain observer.} A passive adversary $\mathcal{O}$ monitors all on-chain transactions and attempts to link them to a single identity or vault. AESP's context-isolated privacy (Section~\ref{sec:privacy}) ensures that transactions using different context strings produce cryptographically independent addresses. From the addresses alone, $\mathcal{O}$ cannot determine whether two ephemeral addresses belong to the same principal without breaking HKDF or compromising the master key. Once an ephemeral address sends funds to the vault, or is funded for fees from it, $\mathcal{O}$ links that address to the vault directly.

\subsubsection{Defense Boundaries}

We state the following defense guarantees and their boundaries explicitly:

\begin{itemize}[leftmargin=*]
\item \textbf{Policy enforcement guarantee:} Under deployment assumption DA1 (mandatory policy mediation), no action violating the active policy can reach the signing layer without first being routed to the ReviewManager. This guarantee is \emph{architectural}: it depends on the host environment enforcing that the cryptographic signing API is not directly accessible to the agent runtime. In deployments where DA1 is not enforced (\eg, the agent process has direct access to signing exports), a compromised agent can bypass the policy gate entirely.

\item \textbf{Key isolation guarantee:} Private keys never leave the Rust module's memory space. In WASM deployments, signing is performed within WASM linear memory; in native FFI deployments, keys remain in the Rust process's address space. Zeroization on drop is enforced in both paths. This guarantee fails if the adversary can modify the compiled binary or exploit a runtime vulnerability in the host environment.

\item \textbf{Sovereignty guarantee:} The human principal $H$ can freeze any agent at any time, after which the Worker signs nothing further for that agent. A freeze does not recall transactions already signed and released, funded escrows, or allowances already granted on the settlement layer; these must be revoked at Layer~1. This guarantee holds as long as the DSE device is not compromised.
\end{itemize}

\subsubsection{Failure Cases}

AESP does not fully mitigate the following three failure modes.

\textbf{Collusion attack.} If both the buyer agent $A_i$ and seller agent $A_j$ are compromised (or their principals collude), they can construct mutually beneficial but policy-violating commitments. AESP's policy gate operates per-agent; cross-agent collusion detection requires a reputation or anomaly-detection layer that is out of AESP's current scope.

\textbf{Approval fatigue.} If an adversary triggers a high volume of review requests (\eg, 50 requests in 30 minutes), the human principal may begin approving requests without careful inspection. AESP's review system supports urgency levels and expiration deadlines but does not currently implement rate-limiting or fatigue detection. We quantify this risk in Section~\ref{sec:eval_limitations}.

\textbf{Time-window boundary exploitation.} An adversary aware of the policy's time window boundaries can queue transactions to execute at the exact moment the window opens, potentially overwhelming the budget tracker before the human can react. AESP's rolling budget limits (daily/weekly/monthly) bound the total, since budget checks reserve atomically (Section~\ref{sec:policy}), but a burst can still spend the whole remaining daily budget within seconds of the window opening.

\subsubsection{Out of Scope}

We do not defend against: (1)~a compromised settlement layer (requires on-chain security guarantees orthogonal to AESP); (2)~a compromised DSE device (if the adversary controls the human's phone, all AESP guarantees are void); (3)~side-channel attacks on the cryptographic module (requires constant-time implementations at the hardware level); or (4)~social engineering of the human principal into approving malicious requests (requires user education, not protocol design).

\section{Identity}
\label{sec:identity}

\subsection{Agent Derivation}

AESP derives agent identities deterministically from the human principal's ACE-GF mnemonic. In REV32 mode, the mnemonic decodes to a 32-byte REV payload $r_H$, and key material is derived through the ACE-GF pipeline:
\begin{equation}
\begin{split}
m_H &= \mathrm{Unseal}_{\mathrm{Argon2id}(p_H, \mathrm{salt}_{\mathrm{pub}})}(r_H), \\
k_H &= \mathrm{HKDF}(m_H, \texttt{"acegf:rev32:kmaster"}), \\
\mathrm{root}_H &= \mathrm{HKDF}(k_H, \texttt{"acegf:identity:root"})
\end{split}
\label{eq:derivation}
\end{equation}
where $p_H$ is the owner's passphrase domain, $\mathrm{salt}_{\mathrm{pub}}$ is the fixed public Argon2id salt (or a deployment-supplied public identifier), and $\mathrm{Unseal}$ is AES-256-GCM-SIV decryption of the 32-byte artifact, which yields the 16-byte identity material $m_H$. In this wallet encoding profile the 32 bytes hold only the 16-byte ciphertext and the 16-byte tag; there is no per-artifact salt field~\cite{acegf2026}. The HKDF notation follows RFC~5869~\cite{rfc5869}; the full Extract-and-Expand decomposition is given in Equation~\ref{eq:hkdf_pipeline} (Section~\ref{sec:rev32}). Agent-specific keys are then derived deterministically from $\mathrm{root}_H$ under agent-scoped derivation context, yielding $(\mathrm{sk}_i, \mathrm{pk}_i)$ for each agent index $i$.
\begin{equation}
\mathrm{id}(A_i) = \mathrm{SHA\text{-}256}(\mathrm{pk}_i)
\end{equation}

The agent is assigned a Decentralized Identifier (DID)~\cite{w3cdid} of the form $\texttt{did:aesp:}\mathrm{id}(A_i)$. This derivation is deterministic: the same ACE-GF mnemonic, passphrase domain, and agent index produce the same agent identity, enabling deterministic recovery.

\subsection{Identity Certificates}

To establish verifiable authorization, the human principal $H$ issues an \emph{identity certificate} for each agent:
\begin{equation}
\begin{split}
\mathrm{cert}(A_i) = \langle \mathrm{ver}, \mathrm{id}(A_i), \mathrm{pk}_i, \mathrm{xid}_H, C, h(\pi), \\
\alpha_{\max}, \mathrm{chains}, t_{\mathrm{created}}, t_{\mathrm{expires}}, \sigma_H \rangle
\end{split}
\end{equation}
where $\mathrm{ver}$ is the certificate version (currently ``1.0''), $\mathrm{xid}_H$ is the human's xidentity public key, $C$ is the set of granted capabilities (\eg, \{payment, negotiation\}), $h(\pi)$ is the SHA-256 hash of the associated policy object, $\alpha_{\max}$ is the maximum autonomous transaction amount per asset, and $\sigma_H$ is the human's Ed25519 signature over the certificate fields.

Verification requires the trusted owner xidentity $\mathrm{xid}_H$. A verifier checks that $\sigma_H$ is a valid Ed25519 signature by $\mathrm{xid}_H$ over the canonical serialization, that the certificate has not expired, and that the claimed capabilities are within the scope of the associated policy.

\subsection{Agent Hierarchy}

AESP supports hierarchical delegation up to a configurable maximum depth $d_{\max} = 5$. The hierarchy forms a rooted tree with the human principal at the root:
\begin{equation}
H \to A_0 \to A_{0,1} \to A_{0,1,2} \to \cdots
\end{equation}

Each agent can only delegate capabilities that are a subset of its own. The \texttt{AgentHierarchyManager} enforces the depth constraint and provides an \emph{escalation chain} function that, given agent $A_i$, returns the ordered sequence $[A_i, \mathrm{parent}(A_i), \ldots, H]$ which is attached to each review request as context. Review requests are always routed to $H$; a parent agent cannot approve a sub-agent's request. A sub-agent is created only when $H$ signs its certificate and policy (Section~\ref{sec:threat}, DA1). The Worker accepts a sub-agent policy only if its capabilities are a subset of the parent's and its limits are reserved out of the parent's remaining budget, and every sub-agent spend is also debited from the parent's budget, so the spending of an agent subtree never exceeds the root agent's limits.

\section{Cryptographic Foundation}
\label{sec:crypto}

AESP uses ACE-GF as its cryptographic execution substrate~\cite{acegf2026}. This layer provides deterministic identity reconstruction, context-isolated derivation, and multi-curve signing capabilities required by AESP. The same core implementation is exposed through WebAssembly and native FFI targets, so browser, server, and mobile deployments share one implementation.

\subsection{Architecture}

The ACE-GF substrate is organized as a layered implementation:

\begin{itemize}[leftmargin=*]
\item \textbf{Core derivation layer}: REV32 reconstruction, HKDF domain separation, and context-isolated per-chain key streams.
\item \textbf{Cryptographic operations layer}: Signing, encryption, and key agreement APIs with curve-specific dispatch.
\item \textbf{Signer layer}: Chain-specific transaction signing for EVM, Solana, Bitcoin, and other supported ecosystems.
\item \textbf{Platform bindings}: WASM and native FFI interfaces that keep a single cryptographic core while adapting to host runtimes.
\item \textbf{Security utilities}: Argon2id-based credential hardening, REV32 artifact handling, and memory zeroization.
\end{itemize}

\subsection{Supported Cryptographic Primitives}

Table~\ref{tab:crypto_primitives} summarizes the cryptographic primitives available in the cryptographic module (all primitives are available across all platform targets unless noted).

\begin{table}[h]
\centering
\caption{Cryptographic Primitives in the ACE-GF Substrate}
\label{tab:crypto_primitives}
\small
\begin{tabular}{@{}lll@{}}
\toprule
\textbf{Category} & \textbf{Primitive} & \textbf{Usage in AESP} \\
\midrule
\multirow{3}{*}{Signatures}
  & Ed25519 & Agent identity, Solana tx \\
  & ECDSA/secp256k1 & EVM tx, EIP-712 signing \\
  & Schnorr/secp256k1 & Bitcoin Taproot tx \\
\addlinespace
Post-Quantum & ML-DSA-44 (FIPS 204) & Future-proof agent certs \\
\addlinespace
Key Agreement & X25519 ECDH & E2EE negotiation msgs \\
\addlinespace
\multirow{3}{*}{Encryption}
  & AES-256-GCM & Negotiation msg encryption \\
  & AES-256-GCM-SIV & Passphrase sealing \\
  & ChaCha20-Poly1305 & Alternative encryption \\
\addlinespace
\multirow{2}{*}{KDF}
  & Argon2id & Passphrase + REV32 $\to$ Kmaster \\
  & HKDF-SHA256 & Context-isolated derivation \\
\addlinespace
Hash & SHA-256 & Identity, commitments \\
\bottomrule
\end{tabular}
\end{table}

\subsection{REV32 Wallet Format and Context-Isolated Derivation}
\label{sec:rev32}

The cryptographic layer uses the REV32 wallet format, which uses identity-root-based context-isolated key generation (rather than HD path extension). The derivation pipeline operates as follows:

\begin{enumerate}[leftmargin=*]
\item \textbf{REV32 recovery.} Decode the ACE-GF mnemonic into the 32-byte REV32 payload (16-byte sealed material and 16-byte AES-GCM-SIV tag, with no metadata).
\item \textbf{Identity-root derivation.} Derive an identity root from the sealed REV32 secret and passphrase domain, then derive chain-specific key material from that root.
\item \textbf{Context extension.} Given a context string $\mathrm{ctx}$, compute a derived key using HKDF:
\begin{equation}
\begin{split}
\mathrm{prk} &= \mathrm{HKDF\text{-}Extract}(\mathrm{salt}, \mathrm{identity\_root}) \\
\mathrm{dk} &= \mathrm{HKDF\text{-}Expand}(\mathrm{prk}, \mathrm{info}, L)
\end{split}
\label{eq:hkdf_pipeline}
\end{equation}
where $\mathrm{info} = \texttt{"ACEGF-REV32-V1-"} \| \mathrm{curve} \| \texttt{":"} \| \mathrm{ctx}$ and $L$ is the key length for the target curve.
\item \textbf{Address generation.} The derived key $\mathrm{dk}$ is used to compute the chain-specific address (Ed25519 pubkey for Solana, Keccak-256 of ECDSA pubkey for EVM, etc.).
\end{enumerate}

This pipeline supports seven address namespaces: Solana, Ethereum/EVM, and Bitcoin (including Taproot) have full signer implementations; Cosmos and Polkadot have address derivation support; ML-DSA-44 provides a post-quantum namespace; and xidentity serves as AESP's native identity namespace.

The property AESP relies on is \emph{context isolation}: given two distinct context strings $\mathrm{ctx}_1 \neq \mathrm{ctx}_2$, the derived keys $\mathrm{dk}_1$ and $\mathrm{dk}_2$ are computationally independent. This property follows from the pseudorandomness of HKDF~\cite{rfc5869}, and AESP's privacy mechanism (Section~\ref{sec:privacy}) depends on it.

The cryptographic module exposes context-isolated derivation through a unified API:
\begin{equation}
\texttt{view\_wallet\_unified\_with\_context}(\mathrm{credential}, \mathrm{ctx}) \to \mathcal{W}
\end{equation}
where $\mathcal{W}$ contains addresses for all supported chains, derived under the given context. The TypeScript SDK calls this function to generate privacy-preserving ephemeral addresses.

\subsection{Key Protection and Memory Safety}

The cryptographic module implements several key protection measures:

\begin{itemize}[leftmargin=*]
\item \textbf{Memory isolation.} All cryptographic operations execute within the Rust module's own memory space: WASM linear memory in browser/Node.js deployments, or the native process heap in C~FFI and Dart~FFI deployments. In either case, the host runtime only sees public keys, addresses, and signed outputs; private keys never cross the binding boundary.
\item \textbf{Zeroization.} All sensitive data structures implement Rust's \texttt{Zeroize} and \texttt{ZeroizeOnDrop} traits, ensuring that key material is overwritten when it leaves scope.
\item \textbf{Argon2id sealing.} When wallet secret material must be persisted (e.g., in browser storage or application keychain), it is protected with AES-256-GCM-SIV under a key derived from the user's passphrase via Argon2id. The default parameters (memory cost = 4~MB, time cost = 3, parallelism = 1) are tuned for constrained WASM environments; native FFI deployments can increase these to match OWASP recommendations.
\item \textbf{No key export.} Neither the WASM nor the C~FFI binding exposes raw private key bytes. The \texttt{sign\_*} and \texttt{derive\_*} functions accept wallet credentials and context, perform operations internally, and return only the result.
\end{itemize}

\subsection{Post-Quantum Readiness}

The cryptographic module includes ML-DSA-44 (Module-Lattice Digital Signature Algorithm, FIPS~204~\cite{fips204}), the NIST-selected post-quantum signature standard. ML-DSA-44 signatures are available as an optional signing scheme for agent identity certificates and transaction authorization. This provides a migration path for AESP deployments that require resistance to quantum computing attacks, without impacting current classical-security deployments that use Ed25519 or secp256k1.

\subsection{EIP-712 Typed Data Signing with Context}

The cryptographic module provides native EIP-712~\cite{eip712} typed data signing with context support:
\begin{equation}
\texttt{evm\_sign\_typed\_data\_with\_context}(\mathrm{credential}, \mathrm{ctx}, \mathrm{typed\_data\_hash}) \to \sigma
\end{equation}

This function derives a secp256k1 keypair under the given context (using the REV32 pipeline), then signs the EIP-712 hash of the typed data. AESP's commitment protocol (Section~\ref{sec:commitment}) uses it to sign each commitment with a key derived from a transaction-specific context string, which prevents correlation by signing key.

\section{Policy Engine}
\label{sec:policy}

\subsection{Policy Structure}

A policy $\pi$ is a tuple:
\begin{equation}
\pi = \langle \mathrm{id}, \mathrm{agentId}, \mathrm{xid}_H, s, \kappa, t_{\mathrm{created}}, t_{\mathrm{expires}} \rangle
\end{equation}
where $s \in \{\texttt{auto\_payment}, \texttt{negotiation}, \texttt{commitment}, \texttt{full}\}$ is the policy scope and $\kappa$ is a conditions record containing the fields shown in Table~\ref{tab:policy_conditions}.

\begin{table}[h]
\centering
\caption{Policy Condition Fields}
\label{tab:policy_conditions}
\small
\begin{tabular}{@{}lll@{}}
\toprule
\textbf{Condition} & \textbf{Type} & \textbf{Semantics} \\
\midrule
\texttt{maxAmountPerTx} & map & Max per-transaction spend \\
\texttt{maxAmountPerDay} & map & Rolling 24-hour limit \\
\texttt{maxAmountPerWeek} & map & Rolling 7-day limit \\
\texttt{maxAmountPerMonth} & map & Calendar-month limit \\
\texttt{allowListAddresses} & string[] & Permitted recipients \\
\texttt{allowListChains} & string[] & Permitted chains \\
\texttt{allowListMethods} & string[] & Permitted methods \\
\texttt{timeWindow} & \{start, end\} & Operating hours (HH:MM) \\
\texttt{minBalanceAfter} & map & Min post-tx balance \\
\texttt{requireReviewFirstPay} & boolean & First pay needs review \\
\bottomrule
\end{tabular}
\end{table}

Amount conditions (type \texttt{map}) are keyed by asset, where an asset is a (chain, token contract) pair, and are expressed in that asset's base units. Checks~1, 7, and~8 compare a request only against the limits of the asset parsed from its payload, and budget totals are kept per asset. A request in an asset for which the policy states no limit does not pass the policy. AESP does not convert between assets, so limits on different assets are independent and do not form a combined cap in any common unit.

Scopes are ranked to enable escalation detection: $\mathrm{rank}(\texttt{auto\_payment}) = 1$, $\mathrm{rank}(\texttt{negotiation}) = 2$, $\mathrm{rank}(\texttt{commitment}) = 3$, $\mathrm{rank}(\texttt{full}) = 10$.

\subsection{Policy Evaluation Algorithm}

When agent $A_i$ requests an action $r$, the policy engine evaluates all active policies for $A_i$ in sequence. For each policy $\pi$, the engine performs eight checks in fixed order:

\begin{enumerate}[leftmargin=*]
\item \textbf{Per-transaction amount.} If $r.\mathrm{amount} > \kappa.\mathrm{maxAmountPerTx}$, reject.
\item \textbf{Time window.} If $\kappa.\mathrm{timeWindow}$ is defined and the current time $t$ falls outside $[\mathrm{start}, \mathrm{end}]$ (handling midnight wrap-around), reject.
\item \textbf{Address allowlist.} If $\kappa.\mathrm{allowListAddresses}$ is non-empty and $r.\mathrm{to} \notin \kappa.\mathrm{allowListAddresses}$, reject.
\item \textbf{Chain allowlist.} If $\kappa.\mathrm{allowListChains}$ is non-empty and $r.\mathrm{chain} \notin \kappa.\mathrm{allowListChains}$, reject.
\item \textbf{Method allowlist.} If $\kappa.\mathrm{allowListMethods}$ is non-empty and $r.\mathrm{method} \notin \kappa.\mathrm{allowListMethods}$, reject.
\item \textbf{First-payment review.} If $\kappa.\mathrm{requireReviewFirstPay} = \mathrm{true}$ and $A_i$ has no prior successful payment under $\pi$, reject.
\item \textbf{Minimum balance.} If $\kappa.\mathrm{minBalanceAfter} > 0$ and the projected post-transaction balance $< \kappa.\mathrm{minBalanceAfter}$, reject.
\item \textbf{Budget limits.} Query the budget tracker for $A_i$'s rolling daily, weekly, and monthly totals in the request's asset, including outstanding reservations. If adding $r.\mathrm{amount}$ would exceed any limit, reject.
\end{enumerate}

If all eight checks pass for any policy, the action is \emph{auto-approved} and the matching policy ID is returned. The Worker performs check~8 and reserves $r.\mathrm{amount}$ in every budget window as one step, with no other request processed between them, so concurrent requests cannot each pass the check against the same remaining budget. Review-approved actions are reserved in the same way. A reservation becomes a debit when the Worker returns a signature and is released only if signing does not complete; it is not released after a signed transaction has left the Worker, because the agent could still broadcast it. If no policy permits the action, it is routed to the review subsystem (Section~\ref{sec:review}).

\textbf{Disjunctive (OR) semantics.} When an agent has multiple active policies, evaluation follows \emph{disjunctive} (OR) semantics: the action is approved if \emph{any} matching policy permits it (Algorithm~\ref{alg:policy}: the loop returns on the first passing policy). This is a deliberate capability-based design~\cite{dennis1966capability}---each policy grants a specific capability (\eg, ``pay up to 100 USDC to vendor~X''), and an agent with multiple policies receives the union of their permissions. Use cases include per-vendor spending limits, coexisting daily-ops and emergency policies with different thresholds, and time-scoped temporary policies that overlap with a permanent base policy. A consequence of OR semantics is that the \emph{most permissive} matching policy determines the outcome; the human principal is responsible for not issuing overly broad policies. This responsibility is enforced by the critical policy change classification (Table~\ref{tab:policy_changes}), which requires human authorization (up to biometric confirmation) for any change that broadens agent permissions.

\begin{algorithm}[h]
\caption{Policy Evaluation}
\label{alg:policy}
\begin{algorithmic}[1]
\REQUIRE Action request $r$, agent $A_i$, active policies $\Pi_{A_i}$
\ENSURE Decision $\in \{\texttt{approved}(\pi), \texttt{review\_required}\}$
\FOR{each $\pi \in \Pi_{A_i}$}
  \STATE $\kappa \leftarrow \pi.\mathrm{conditions}$
  \IF{$r.\mathrm{amount} > \kappa.\mathrm{maxAmountPerTx}$}
    \STATE \textbf{continue}
  \ENDIF
  \IF{$\neg \mathrm{inTimeWindow}(t, \kappa.\mathrm{timeWindow})$}
    \STATE \textbf{continue}
  \ENDIF
  \IF{$\kappa.\mathrm{allowListAddresses} \neq \emptyset \wedge r.\mathrm{to} \notin \kappa.\mathrm{allowListAddresses}$}
    \STATE \textbf{continue}
  \ENDIF
  \IF{$\kappa.\mathrm{allowListChains} \neq \emptyset \wedge r.\mathrm{chain} \notin \kappa.\mathrm{allowListChains}$}
    \STATE \textbf{continue}
  \ENDIF
  \IF{$\kappa.\mathrm{allowListMethods} \neq \emptyset \wedge r.\mathrm{method} \notin \kappa.\mathrm{allowListMethods}$}
    \STATE \textbf{continue}
  \ENDIF
  \IF{$\kappa.\mathrm{requireReviewFirstPay} \wedge \neg \mathrm{hasPriorPayment}(A_i, \pi)$}
    \STATE \textbf{continue}
  \ENDIF
  \IF{$\mathrm{projectedBalance}(r) < \kappa.\mathrm{minBalanceAfter}$}
    \STATE \textbf{continue}
  \ENDIF
  \IF{$\neg \mathrm{reserveBudget}(A_i, r.\mathrm{asset}, r.\mathrm{amount})$}
    \STATE \textbf{continue}
  \ENDIF
  \RETURN $\texttt{approved}(\pi)$
\ENDFOR
\RETURN $\texttt{review\_required}$
\end{algorithmic}
\end{algorithm}

\subsection{Critical Policy Change Classification}

When a policy is modified, AESP classifies the change to determine the required approval level. Eight change types are recognized, as shown in Table~\ref{tab:policy_changes}.

\begin{table}[h]
\centering
\caption{Critical Policy Change Classification}
\label{tab:policy_changes}
\small
\begin{tabular}{@{}lll@{}}
\toprule
\textbf{Change Type} & \textbf{Condition} & \textbf{Approval} \\
\midrule
\texttt{budget\_increase} & Any max* increased & biometric \\
\texttt{scope\_escalation} & rank(new) $>$ rank(old) & biometric \\
\texttt{addr\_remove\_all} & Allowlist cleared & biometric \\
\texttt{addr\_add} & New address added & review \\
\texttt{time\_window\_remove} & Time restriction removed & review \\
\texttt{min\_balance\_lower} & Min balance decreased & review \\
\texttt{first\_pay\_disable} & First-pay review off & review \\
\texttt{expiration\_extend} & Expiration extended & review \\
\bottomrule
\end{tabular}
\end{table}

Changes in the \emph{biometric} set require the human principal to confirm via biometric authentication (\eg, Face~ID, fingerprint) on their mobile device. Changes in the \emph{review} set require explicit confirmation through the review interface. In both cases the Worker applies the change only when it arrives signed by $\mathrm{xid}_H$ (DA1).

\section{Negotiation}
\label{sec:negotiation}

\subsection{State Machine}

Agent-to-agent negotiation in AESP is modeled as a finite state machine (FSM) with eight states and thirteen transitions.

\textbf{States:} $S = \{s_0, s_1, \ldots, s_7\}$ as defined in Table~\ref{tab:negotiation_states}.

\begin{table}[h]
\centering
\caption{Negotiation FSM States}
\label{tab:negotiation_states}
\small
\begin{tabular}{@{}cll@{}}
\toprule
\textbf{State} & \textbf{Name} & \textbf{Description} \\
\midrule
$s_0$ & \texttt{initial} & Session created, no messages \\
$s_1$ & \texttt{offer\_sent} & Initiator sent an offer \\
$s_2$ & \texttt{offer\_received} & Responder received an offer \\
$s_3$ & \texttt{countering} & Active counter-offer exchange \\
$s_4$ & \texttt{accepted} & Both parties agree \\
$s_5$ & \texttt{rejected} & Negotiation terminated \\
$s_6$ & \texttt{committed} & EIP-712 commitment attached \\
$s_7$ & \texttt{disputed} & Dispute filed post-commitment \\
\bottomrule
\end{tabular}
\end{table}

\textbf{Transitions:} The valid transition function $\delta: S \times M \to S$ is defined by thirteen tuples:
\begin{equation}
\begin{split}
\delta = \{ & (s_0, \texttt{offer}) \to s_1, \; (s_0, \texttt{offer\_recv}) \to s_2, \\
& (s_1, \texttt{counter}) \to s_3, \; (s_1, \texttt{accept}) \to s_4, \\
& (s_1, \texttt{reject}) \to s_5, \; (s_2, \texttt{counter}) \to s_3, \\
& (s_2, \texttt{accept}) \to s_4, \; (s_2, \texttt{reject}) \to s_5, \\
& (s_3, \texttt{counter}) \to s_3, \; (s_3, \texttt{accept}) \to s_4, \\
& (s_3, \texttt{reject}) \to s_5, \; (s_4, \texttt{commit}) \to s_6, \\
& (s_6, \texttt{dispute}) \to s_7 \}
\end{split}
\end{equation}

Note that $s_3 \to s_3$ (counter $\to$ counter) is a self-loop, allowing multiple rounds of negotiation. Sessions are bounded by a maximum round count (default: 10) and a time-to-live (default: 24 hours). The current FSM enforces only this session-level TTL; per-state timeouts (\eg, an accepted-state commitment deadline shorter than 24 hours) are delegated to the application layer.

\subsection{Message Format and Encryption}

Negotiation messages are typed as one of four message types: \texttt{negotiation\_offer}, \texttt{negotiation\_counter}, \texttt{negotiation\_accept}, and \texttt{negotiation\_reject}. When transmitted between agents, messages are end-to-end encrypted using X25519 ECDH key agreement (performed by the cryptographic module) to derive a shared secret, followed by AES-256-GCM symmetric encryption. The encrypted envelope includes a version identifier, algorithm descriptor, base64-encoded ciphertext, a unique message ID for replay protection, and a timestamp.

The protocol is transport-agnostic: encrypted messages can be carried over MCP tool calls, HTTP, WebSockets, or any other channel. The \texttt{NegotiationProtocol} class accepts a pluggable \texttt{MessageSender} interface for transport abstraction.

\subsection{Agreement Hash}

When both parties accept, an \emph{agreement hash} is computed:
\begin{equation}
h_{\mathrm{agree}} = \mathrm{SHA\text{-}256}(\mathrm{JSON}(\mathrm{lastRound.payload}))
\end{equation}
where \texttt{lastRound.payload} is the most recently exchanged offer or counter-offer. This hash is included in the acceptance message and can be bound into the EIP-712 commitment (Section~\ref{sec:commitment}) to cryptographically link the negotiation outcome to the payment commitment.

\section{Commitment}
\label{sec:commitment}

\subsection{EIP-712 Structure}

AESP uses EIP-712~\cite{eip712} typed structured data for agent-to-agent payment commitments. The domain separator is:
\begin{multline}
D = \langle \texttt{name}: \text{``YalletAgentCommitment''}, \\
\texttt{version}: \text{``1''}, \; \texttt{chainId}: c \rangle
\end{multline}

The commitment type contains nine fields: \texttt{buyerAgent} (address), \texttt{sellerAgent} (address), \texttt{item} (string), \texttt{price} (uint256), \texttt{currency} (address), \texttt{deliveryDeadline} (uint256), \texttt{arbitrator} (address), \texttt{escrowRequired} (bool), and \texttt{nonce} (uint256, from \texttt{crypto.getRandomValues}).

\subsection{Dual-Signing Flow}

A commitment passes through the following lifecycle:
\begin{equation}
\begin{split}
\texttt{draft} \to \texttt{proposed} \to \texttt{buyer\_signed} \to \texttt{fully\_signed} \\
\to \texttt{escrowed} \to \texttt{delivered} \to \texttt{completed}
\end{split}
\end{equation}
with alternative transitions to \texttt{disputed} (from escrowed or delivered) and \texttt{cancelled} (from any pre-escrowed state). The commitment hash is:
\begin{equation}
h_c = \mathrm{SHA\text{-}256}(\mathrm{JSON}(\langle D, V \rangle))
\end{equation}
where $V$ is the commitment value record and $\mathrm{JSON}(\cdot)$ denotes deterministic JSON serialization with lexicographically sorted keys (matching the SDK's \texttt{JSON.stringify} with sorted-key canonicalization). Both buyer and seller compute $h_c$ independently over the same canonical form and use it as the record identifier. Each party signs the EIP-712 digest $\mathrm{keccak256}(\texttt{0x1901} \| \mathrm{domainSeparator}(D) \| \mathrm{hashStruct}(V))$, which an escrow contract can verify on chain; the Worker computes this digest from the parsed typed data and evaluates the commitment's \texttt{price} and \texttt{currency} against policy as a payment before signing. The \texttt{fully\_signed} state is reached only when both signatures are present. The record identifier $h_c$ is a local index and is not signed; the signed object is the EIP-712 digest, which a verifier recomputes from $(D, V)$, so integrity of a commitment rests on the two signatures over that digest, not on $h_c$.

When context-isolated signing is enabled, each party uses the cryptographic module's \texttt{evm\_sign\_typed\_data\_with\_context} function with a commitment-specific context string, ensuring that the signing key is unique to this transaction.

\subsection{Settlement Integration}

AESP does not implement escrow or fund transfers. It produces a fully-signed commitment record containing the EIP-712 structure, both signatures, and metadata (escrow transaction hash, delivery confirmation hash, release transaction hash). The settlement layer consumes this record to fund escrow, verify delivery, and release payment.

\section{Review (Human-in-the-Loop)}
\label{sec:review}

\subsection{Review Queue}

When the policy engine rejects an action (Section~\ref{sec:policy}), AESP creates a \emph{review request} and places it in a priority queue. Each request includes the original action, the agent ID, the policy violation reason, an urgency level (low, normal, high, critical), and an expiration deadline (default: 30 minutes).

The review subsystem returns a promise that resolves when the human submits a response, signed by $\mathrm{xid}_H$ from the DSE and verified by the Worker before any signing, (approve, reject, or modify) or rejects when the deadline expires. This design allows the calling agent to \texttt{await} the review outcome without polling.

\subsection{Escalation Tiers}

AESP defines three escalation tiers (Table~\ref{tab:escalation}).

\begin{table}[h]
\centering
\caption{Escalation Tiers}
\label{tab:escalation}
\small
\begin{tabular}{@{}lll@{}}
\toprule
\textbf{Tier} & \textbf{Trigger} & \textbf{Required Action} \\
\midrule
Automatic & All policy checks pass & No human involvement \\
Review & Policy check fails & Human confirms via UI \\
Biometric & Critical policy change & Human biometric auth \\
\bottomrule
\end{tabular}
\end{table}

This tiered model follows the progressive validation of~\cite{cheng2026threepillar}; the biometric tier requires physical presence at the device.

\subsection{Emergency Freeze}

The human principal can freeze any agent at any time by issuing a freeze command through the DSE. When an agent is frozen: (1)~all pending review requests for that agent are immediately rejected; (2)~any new action is blocked with an \texttt{AGENT\_FROZEN} error; (3)~the freeze status is persisted and survives restarts. Freezing is the last-resort control.

\subsection{Event System}

The review system emits six event types: \texttt{request\_created}, \texttt{request\_approved}, \texttt{request\_rejected}, \texttt{request\_modified}, \texttt{request\_expired}, and \texttt{request\_cancelled}. External systems (\eg, mobile notification services, audit loggers) can subscribe to these events.

\section{Privacy}
\label{sec:privacy}

\subsection{Problem Statement}

On public blockchains, all transactions are visible to any observer $\mathcal{O}$. If an agent uses the same address for multiple transactions, $\mathcal{O}$ can trivially link them. Even with different addresses, timing patterns, amounts, and counterparties may allow statistical correlation. AESP's privacy subsystem addresses the \emph{transaction unlinkability} problem: given two transactions $\mathrm{tx}_1$ and $\mathrm{tx}_2$ produced by AESP, an observer $\mathcal{O}$ who does not possess the master key should be unable to determine whether they originate from the same vault or principal.

\subsection{Context-Isolated Address Derivation}

AESP derives ephemeral addresses using HKDF~\cite{rfc5869} with a \emph{context string} that encodes the transaction's identity context. The context string is constructed by sorting key-value segments alphabetically and joining with colons:
\begin{equation}
\mathrm{ctx} = \mathrm{sort}(\{ \texttt{agent:}a, \; \texttt{dir:}d, \; \texttt{seq:}n, \; \texttt{tx:}t \}) \| \texttt{':'}
\end{equation}
where $a$ is the agent ID, $d \in \{\texttt{inbound}, \texttt{outbound}\}$, $n$ is a sequence counter, and $t$ is a unique transaction UUID. The sorting ensures deterministic context string generation.

The context string is passed to the cryptographic module's REV32 derivation pipeline (Section~\ref{sec:rev32}), which applies HKDF with:
\begin{equation}
\mathrm{info} = \texttt{"ACEGF-REV32-V1-"} \| \mathrm{curve} \| \texttt{":"} \| \mathrm{ctx}
\end{equation}

\subsection{Privacy Levels}

AESP supports three privacy levels:

\begin{itemize}[leftmargin=*]
\item \textbf{Transparent} ($\ell_0$): The agent uses its main vault address directly. No privacy protection.
\item \textbf{Basic} ($\ell_1$): One shared address per agent-chain-direction triple. Context: $\{\texttt{agent:}a, \texttt{dir:}d, \texttt{mode:basic}\}$.
\item \textbf{Isolated} ($\ell_2$): A unique address per transaction. Context includes a unique transaction ID, ensuring one-time use.
\end{itemize}

\subsection{Address Pool}

To avoid derivation latency at transaction time, AESP pre-derives a pool of ephemeral addresses for each agent-chain-direction triple. The default pool size is~5. When an address is claimed, the pool asynchronously replenishes. Pool addresses use the context segment \texttt{pool:pre} and an incrementing sequence counter; when claimed, they are mapped to the specific transaction context.

\subsection{Consolidation}

Ephemeral addresses accumulate funds that must eventually be consolidated back into the vault. Any transfer from an ephemeral address to the vault links that address to the vault on chain, and a multi-input batch on a UTXO chain also links its inputs to one another; the same holds when fees for an account-based chain are funded from the vault. AESP's \texttt{ConsolidationScheduler} does not remove these links. Its three techniques only reduce timing and ordering correlation:

\textbf{Timing jitter.} The base consolidation interval (default: 4~hours) is perturbed by a configurable jitter ratio (default: ${\pm}30\%$):
\begin{equation}
t_{\mathrm{actual}} = t_{\mathrm{base}} \cdot (1 - \rho + 2\rho \cdot r), \quad r \sim \mathrm{Uniform}(0,1)
\end{equation}
where $\rho$ is the jitter ratio.

\textbf{Address shuffle.} Before consolidation, the set of ephemeral addresses is permuted using the Fisher-Yates algorithm:
\begin{equation}
\text{for } i = n{-}1 \text{ downto } 1: \; j \leftarrow \mathrm{random}[0, i]; \; \mathrm{swap}(a_i, a_j)
\end{equation}
ensuring that consolidation order reveals no information about derivation order.

\textbf{Batched execution.} Addresses are consolidated in batches of configurable size (default: 5), with random inter-batch delays drawn uniformly from $[10, 60]$ minutes.

\subsection{Audit Tags}

Despite unlinkability guarantees, the asset owner must reconstruct transaction history for auditing. AESP's \texttt{ContextTagManager} creates an encrypted \emph{context tag} for each ephemeral transaction, recording agent ID, policy ID, commitment ID, ephemeral address, and metadata. Tags are stored locally and optionally archived to permanent storage (\eg, Arweave) in encrypted form.

Archiving supports three batching strategies: \emph{immediate} (each tag archived individually), \emph{time window} (batched after a configurable window, default: 5~minutes), and \emph{count threshold} (batched when count reaches a threshold, default: 50).

\subsection{Privacy Design Rationale}

The privacy design rests on the following argument. At the \emph{address derivation layer}, HKDF~\cite{rfc5869} with distinct info strings produces outputs that are computationally indistinguishable from independent random values, assuming the input keying material has sufficient min-entropy (in the REV32 wallet profile, the identity root is expanded from 128 bits of source entropy). Addresses derived from different context strings are therefore cryptographically unrelated, and an observer who sees only on-chain addresses cannot determine whether they share a common master key without breaking HKDF's pseudorandomness.

However, address-level unlinkability does not imply full transaction unlinkability in practice. Side channels such as consolidation timing, transaction amounts, and counterparty patterns can leak correlation information even when addresses are independent. AESP's consolidation countermeasures (timing jitter, Fisher-Yates shuffle, batched execution) are designed to reduce these side channels, but we do not claim formal unlinkability against all adversaries. Address-level unlinkability holds only for ephemeral addresses that have not been consolidated to, or funded from, the vault. Section~\ref{sec:eval_privacy} describes the planned empirical evaluation of residual linkability under a heuristic clustering adversary. A formal privacy analysis in a simulation-based framework (\eg, the Transaction DAG model of Wicht~\etal{}~\cite{wicht2024privacy}) is left to future work.

\section{Implementation}
\label{sec:implementation}

\subsection{SDK Architecture}

AESP is implemented as a TypeScript library (\texttt{@yallet/aesp}) targeting ES2022 with ECMAScript module (ESM) output. The implementation is organized into ten modules, each available as a separate entry point for tree-shaking (Table~\ref{tab:modules}).

\begin{table}[h]
\centering
\caption{AESP SDK Modules}
\label{tab:modules}
\small
\begin{tabular}{@{}ll@{}}
\toprule
\textbf{Module} & \textbf{Primary Exports} \\
\midrule
\texttt{types} & All shared type definitions \\
\texttt{crypto} & Signing, encryption, hashing \\
\texttt{identity} & \texttt{deriveAgentIdentity}, certificates \\
\texttt{policy} & \texttt{PolicyEngine}, \texttt{BudgetTracker} \\
\texttt{negotiation} & \texttt{NegotiationStateMachine}, protocol \\
\texttt{commitment} & \texttt{CommitmentBuilder} \\
\texttt{review} & \texttt{ReviewManager} \\
\texttt{mcp} & \texttt{MCPServer}, tool definitions \\
\texttt{a2a} & \texttt{AgentCardBuilder} \\
\texttt{privacy} & \texttt{AddressPoolManager}, consolidation \\
\bottomrule
\end{tabular}
\end{table}

\subsection{Cryptographic Backend}

All cryptographic operations are performed by the Rust module described in Section~\ref{sec:crypto}. The TypeScript SDK interacts with the module through a thin adapter layer that handles serialization/deserialization---via \texttt{wasm-bindgen} in browser/Node.js deployments or via C~FFI bindings in server-side and mobile deployments. For SHA-256 hashing only, a Web Crypto API fallback is provided when the cryptographic module is unavailable.

\subsection{Storage Abstraction}

All stateful modules accept a pluggable \texttt{StorageAdapter} interface:
\begin{verbatim}
interface StorageAdapter {
  getItem(key: string): Promise<string | null>;
  setItem(key: string, value: string):
    Promise<void>;
}
\end{verbatim}
Tests use an in-memory adapter and servers a database. For policy state the adapter must meet DA1: it runs inside the Worker and writes to storage the agent runtime cannot write, so \texttt{localStorage} in the agent's origin is not used for it.

\subsection{Testing}

The implementation includes 208 tests across nine test suites (covering all ten modules), executed with Vitest. Test coverage spans identity derivation, policy evaluation (including all eight check types and all eight critical change classifications), negotiation FSM transitions, commitment lifecycle, review queue and freeze mechanics, MCP tool validation, A2A card generation, cryptographic operations, and privacy features (address derivation, consolidation jitter, Fisher-Yates shuffle, audit batching).

\section{Interoperability}
\label{sec:interop}

\subsection{Model Context Protocol (MCP)}

AESP defines eight MCP~\cite{mcp2025} tools that expose economic operations to any MCP-compatible AI framework (Table~\ref{tab:mcp_tools}).\footnote{The \texttt{yault\_} prefix reflects the reference settlement layer integration (Yault). Deployments targeting other settlement layers may rename these tools while preserving the same JSON Schema interfaces.}

\begin{table}[h]
\centering
\caption{MCP Tool Definitions}
\label{tab:mcp_tools}
\small
\begin{tabular}{@{}ll@{}}
\toprule
\textbf{Tool} & \textbf{Description} \\
\midrule
\texttt{yault\_check\_balance} & Query agent account balance \\
\texttt{yault\_deposit} & Deposit to ERC-4626 vault \\
\texttt{yault\_redeem} & Redeem vault shares \\
\texttt{yault\_create\_allowance} & Create payment allowance \\
\texttt{yault\_cancel\_allowance} & Cancel an allowance \\
\texttt{yault\_file\_dispute} & File dispute with evidence hash \\
\texttt{yault\_check\_budget} & Query remaining budget \\
\texttt{yault\_list\_agents} & List sub-agents with status \\
\bottomrule
\end{tabular}
\end{table}

Each tool includes a JSON Schema definition for argument validation. The \texttt{MCPServer} class provides a transport layer supporting both stdio and server-sent events (SSE).

\subsection{Agent-to-Agent (A2A) Protocol}

AESP generates A2A agent cards~\cite{a2a2025} from agent configuration. Each card advertises the agent's capabilities (payment, negotiation, data query, commitment, delegation, arbitration), supported input/output modes, authentication scheme (Ed25519), and endpoint URL. An agent running on a different platform can thus discover an AESP-managed agent's economic capabilities and initiate interaction.

\section{Evaluation Methodology}
\label{sec:evaluation}

This section specifies the evaluation methodology for validating the two hypotheses stated in Section~\ref{sec:intro}. All experiments execute at the SDK layer using the AESP TypeScript library and the cryptographic module (WASM binding for the primary benchmark environment); no on-chain transactions are required. The experiment framework, runner scripts, and reproduction instructions are maintained alongside the source code.\footnote{\url{https://github.com/ya-xyz/aesp}} This version focuses on methodology and experimental design; empirical result tables are deferred to a dedicated evaluation report.

\subsection{Evaluation Design}

\subsubsection{Request Corpus}

We define a synthetic request corpus of 1,500 action requests:

\begin{itemize}[leftmargin=*]
\item \textbf{Attack set} ($n = 1{,}000$): Requests representing unauthorized agent activity, divided into two categories. \emph{Single-condition violations} ($n \approx 950$) each violate at least one individual policy condition, stratified across: over-limit amounts, out-of-window timing, disallowed addresses, disallowed chains, disallowed methods, first-payment violations, minimum-balance violations, and budget-exceeding requests. \emph{Aggregate violations} ($n \approx 50$) are individually policy-compliant---each passes all eight checks in isolation---but represent unauthorized behavior detectable only through cross-request context (\eg, rapid sequences of compliant transactions that stay within budget but exhaust the daily allowance in minutes).

\item \textbf{Legitimate set} ($n = 500$): Requests that comply with all policy conditions, covering normal transaction patterns across varied amounts (0.01--100 units), permitted addresses, permitted chains, and within-window timing. Amounts are drawn from a log-normal distribution to simulate realistic spending patterns.
\end{itemize}

Each request is a structured \texttt{PolicyCheckRequest} object with fields for amount, recipient address, chain ID, method, timestamp, and balance context. The reference policy uses representative conditions, all in a single reference asset: \texttt{maxAmountPerTx}~=~100, \texttt{maxAmountPerDay}~=~500, \texttt{maxAmountPerWeek}~=~2{,}000, \texttt{maxAmountPerMonth}~=~5{,}000, \texttt{timeWindow}~=~09:00--21:00, \texttt{minBalanceAfter}~=~10, \texttt{requireReviewFirstPay}~=~true, with 10 addresses in the allowlist, 3 chains, and 4 methods.

\subsubsection{Baselines}

We compare AESP against four baselines of increasing restrictiveness. Because there is no standardized public benchmark suite or directly comparable deployed protocol that exposes the same eight-check gate with tiered human escalation, these baselines are designed as \emph{mechanism-isolation controls}: each step adds one layer of governance complexity, allowing us to attribute security and latency changes to specific protocol components rather than to implementation differences across unrelated systems.

\begin{itemize}[leftmargin=*]
\item \textbf{B0 (Unconstrained):} No policy checks. All requests are auto-approved. This represents a raw wallet with no agent restrictions.
\item \textbf{B1 (Amount-only):} Only the per-transaction amount check (\texttt{maxAmountPerTx}). Simulates a simple spending cap.
\item \textbf{B2 (Two-check):} Amount check + time window check. Simulates a spending cap with operating hours.
\item \textbf{B3 (Full gate, no escalation):} All eight policy checks are applied, but rejected actions are silently dropped (no human review, no escalation). Simulates a strict automated filter without human-in-the-loop.
\item \textbf{AESP (Full):} All eight policy checks + three-tier escalation. Rejected actions are routed to the ReviewManager; auto-approval, explicit review, and biometric tiers are active.
\end{itemize}

\subsubsection{Metrics}

For H1 (Security), we report auto-block performance as the primary security outcome and treat escalation as operational load:
\begin{itemize}[leftmargin=*]
\item \textbf{Auto-blocked rate:} Fraction of attack requests deterministically rejected by the policy gate without human involvement. This is the \emph{guaranteed} defense: auto-blocked requests cannot proceed regardless of human behavior.
\item \textbf{Escalation load rate:} Fraction of attack requests routed to the human principal via the ReviewManager. This is an operational burden metric, not counted as deterministic security success.
\item \textbf{False positive rate (FPR):} Fraction of legitimate requests auto-blocked or escalated.
\item \textbf{Per-check attribution:} Which of the eight checks triggered auto-blocking, used in the ablation study.
\end{itemize}

For H2 (Efficiency), we measure:
\begin{itemize}[leftmargin=*]
\item \textbf{Policy evaluation latency:} Time for the 8-check sequence (\texttt{evaluatePolicy()}).
\item \textbf{Signing latency:} Ed25519 and secp256k1 signing time (via the cryptographic module).
\item \textbf{Key derivation latency:} HKDF context-isolated derivation time.
\item \textbf{End-to-end overhead:} Total latency from action request to signed authorization (policy check + signing + derivation).
\end{itemize}

All latency measurements will use \texttt{performance.now()} with 100 warm-up iterations discarded and $\geq$1,000 measured iterations. Results will be reported as median with interquartile range (IQR) across $\geq$5 independent trials. The primary benchmark environment is a modern Node.js runtime (e.g., v25) with the WASM binding loaded via \texttt{wasm-bindgen}; comparative measurements using the native C~FFI binding will be reported where applicable.

\subsubsection{Ablation Design}

To isolate the contribution of each policy check, we remove one check at a time from the full eight-check gate (B3 configuration, without escalation) and re-evaluate the auto-blocked rate. The expected output is a delta table showing the marginal contribution of each check to overall security coverage.

\subsection{Planned Analyses}

\subsubsection{Security Coverage (H1)}

The baseline comparison will report the auto-blocked rate, escalation load rate, and FPR for each of the five configurations (B0--B3 and AESP Full). The hypothesis thresholds are an auto-blocked rate of at least 90\% and an FPR of at most 5\%. We expect the eight-check gate to auto-block the vast majority of single-condition violations, while escalation load will be concentrated in aggregate violations that pass individual checks but exhibit suspicious cross-request patterns. The ablation will quantify the marginal contribution of each check.

\textbf{Interpreting escalation.} Escalation is not treated as automatic interception. If the human approves all escalated requests without inspection, effective protection equals the auto-blocked rate alone. We therefore report escalation only as review load and analyze its usability implications separately (Section~\ref{sec:eval_limitations}).

\subsubsection{Latency Overhead (H2)}

We will measure per-operation latency for: 8-check policy evaluation, budget tracker query, Ed25519 signing, secp256k1 signing, EIP-712 typed data signing, HKDF context derivation, SHA-256 hashing (cryptographic module vs.\ Web Crypto fallback), and composite end-to-end paths. All cryptographic operations are benchmarked via the WASM binding as the primary target; native C~FFI measurements will be reported as a secondary comparison. The hypothesis threshold is 200\,ms median end-to-end overhead; we expect actual overhead to be substantially lower given that individual cryptographic operations typically complete in single-digit to low-double-digit milliseconds.

\subsection{Transaction Unlinkability}
\label{sec:eval_privacy}

To assess the privacy subsystem, we will generate $N = 1{,}000$ transactions across multiple agents, each using context-isolated addresses (privacy level $\ell_2$), and apply a simulated address-clustering heuristic (common-input-ownership and temporal proximity) to the resulting on-chain footprint.

The experiment will compare three configurations: (1)~no consolidation countermeasures (baseline), (2)~timing jitter ($\pm$30\%) alone, and (3)~jitter + Fisher-Yates shuffle + batched consolidation (batch size 5, inter-batch delay 10--60\,min). We expect progressive reduction in timing-based linkability across these configurations. Every consolidated address remains linked to the vault by its transfer destination in all three configurations, so the experiment reports linkability among addresses not yet consolidated separately from consolidated ones.

\subsection{Anticipated Threats to Validity}
\label{sec:eval_limitations}

\textbf{Construct validity.} The attack corpus is synthetic; real-world attack distributions may differ. We mitigate this by stratifying across all eight check types and including compound attacks that violate multiple conditions simultaneously.

\textbf{External validity.} Latency measurements will be platform-specific. WASM performance varies across runtimes (browsers, Deno, server-side Node.js), and native FFI performance will differ from WASM results. Browser-based deployments may show 1.5--2$\times$ higher overhead due to JIT warmup, while native C~FFI deployments are expected to be faster.

\textbf{Approval fatigue.} The evaluation assumes the human principal responds optimally to review requests. In practice, high review volumes may degrade human decision quality. Addressing approval fatigue through rate-limiting and intelligent batching is future work (see Section~\ref{sec:discussion}).

\section{Case Studies}
\label{sec:casestudies}

We present three case studies, each exercising a different combination of protocol features. Each case study is implemented as an end-to-end scenario using the AESP SDK.

\subsection{Three-Party Grocery Delivery}

This scenario models a consumer agent, a grocery store agent, and a delivery agent coordinating a grocery delivery transaction. The consumer agent negotiates prices with the grocery store agent through the negotiation FSM (Section~\ref{sec:negotiation}), forms an EIP-712 commitment (Section~\ref{sec:commitment}), and coordinates delivery payment. Key protocol features exercised:

\begin{itemize}[leftmargin=*]
\item Multi-agent identity derivation (three agents from separate principals)
\item Cross-agent negotiation with offer/counter/accept flow
\item Dual-signed commitment consumed by an external escrow contract
\item Policy enforcement (spending limits, chain restrictions)
\item Human review for the initial high-value grocery order
\end{itemize}

\subsection{Autonomous Cloud Resource Optimizer}

This scenario models an AI agent that autonomously manages cloud infrastructure spending. The agent monitors resource utilization and negotiates spot instance prices, subject to human-defined budget policies. Key protocol features exercised:

\begin{itemize}[leftmargin=*]
\item Policy engine with budget tracking (daily/weekly/monthly limits)
\item Critical policy change classification (budget increase requires biometric)
\item Emergency freeze/unfreeze cycle
\item Rolling budget window management
\end{itemize}

\subsection{Privacy-Preserving NFT Hunter}

This scenario models an agent that scouts and acquires NFTs across multiple marketplaces while preserving the buyer's identity privacy. Key protocol features exercised:

\begin{itemize}[leftmargin=*]
\item All three privacy levels (transparent, basic, isolated)
\item Address pool replenishment and claiming
\item Batched consolidation with Fisher-Yates shuffle and timing jitter
\item Context-isolated address derivation via the cryptographic module
\item Encrypted audit tag archiving
\end{itemize}

The grocery scenario stresses multi-party coordination, the cloud optimizer policy enforcement and human oversight, and the NFT hunter privacy.

\section{Discussion and Limitations}
\label{sec:discussion}

\textbf{Evaluation status.} Section~\ref{sec:evaluation} specifies the evaluation methodology (request corpus, baselines, metrics, and ablation design) for both hypotheses; empirical results are not yet reported. The framework separates deterministic protection (auto-block rate) from operational burden (escalation load) and measures end-to-end latency overhead of policy evaluation, cryptographic signing, and key derivation.

\textbf{Settlement layer dependency.} AESP defines the protocol and produces signed authorizations, but does not implement on-chain execution. End-to-end operation requires a conforming settlement layer. The current implementation is designed for integration with Yault but is architecturally settlement-agnostic.

\textbf{Privacy limitations and residual leakage.} Context-isolated privacy depends on a wallet that supports HKDF-based context derivation in REV32 mode. As described in Section~\ref{sec:eval_privacy}, consolidation countermeasures (timing jitter, Fisher-Yates shuffle, batched consolidation) are designed to reduce address linkability, but they do not hide the destination of a consolidation. Consolidating to the vault links each consolidated ephemeral address to it, and multi-input batches link their inputs to each other. Amount-based correlation adds further leakage. The unlinkability provided by the current design therefore lasts only until funds are consolidated. Consolidating to fresh addresses, mixing, or spending directly from ephemeral addresses could reduce this, but none is currently implemented.

\textbf{Approval fatigue.} The \texttt{requireReviewFirstPay} check and budget-boundary proximity are expected to produce a non-trivial false positive rate. Under sustained high-throughput operation, this could generate review fatigue. AESP currently supports urgency levels and expiration deadlines but does not implement rate-limiting, intelligent batching of review requests, or adaptive thresholds that learn from human approval patterns. Addressing this is a priority for future work.

\textbf{Binding performance variance.} Cryptographic latency varies by binding target. WASM execution in browsers may show 1.5--2$\times$ higher overhead than Node.js due to JIT compilation warmup and stricter memory constraints. Server-side and mobile deployments using the native C~FFI or Dart~FFI bindings bypass the WASM layer entirely and are expected to match native Rust performance.

\textbf{Post-quantum migration.} While ML-DSA-44 is available in the cryptographic module (across all binding targets), a complete post-quantum migration requires post-quantum key agreement (replacing X25519) and post-quantum encryption (replacing AES-256-GCM with a post-quantum KEM + symmetric scheme). The current post-quantum support covers signatures only.

\textbf{Scalability.} The review queue introduces latency for out-of-policy actions. In a high-throughput agent economy, the 30-minute default review deadline may be too slow. Adaptive deadline policies and batch review interfaces are areas for improvement.

\textbf{Formal verification.} While the negotiation FSM and policy evaluation sequence are specified precisely, the implementation has not been formally verified. Applying model checking to the state machine and policy evaluation logic is a natural next step.

\textbf{Regulatory compliance.} AESP's policy engine and audit trail support compliance requirements (spending limits, allowlists, audit records), but jurisdiction-specific regulations (\eg, money transmitter classification, MiCA) require analysis at the settlement and platform layers.

\section{Conclusion}
\label{sec:conclusion}

We have presented AESP, a protocol for enabling AI agents to participate in economic transactions while ensuring that economic sovereignty remains with the human principal. Under deployment assumption DA1, the protocol enforces the invariant that agents are \emph{economically capable but never economically sovereign} through a combination of deterministic policy evaluation, tiered human-in-the-loop review, dual-signed EIP-712 commitments, HKDF-based context-isolated privacy, and the ACE-GF cryptographic module with WASM and native FFI bindings.

We specify an evaluation methodology with four baselines of increasing restrictiveness, a per-check ablation design, and a transaction unlinkability experiment, targeting two falsifiable hypotheses: (H1) the eight-check gate auto-blocks $\geq$90\% of unauthorized transactions with $\leq$5\% false positive rate, while escalation load is reported separately; and (H2) end-to-end latency overhead remains below 200\,ms.

AESP implements requirements identified in recent conceptual work, notably Google DeepMind's Intelligent AI Delegation framework~\cite{tomasev2026delegation}. The accompanying open-source TypeScript SDK (208 tests, ten modules)\footnote{\url{https://github.com/ya-xyz/aesp}} provides interoperability with MCP and A2A.

\bibliographystyle{IEEEtran}

\appendices

\section{Protocol Constants}
\label{app:constants}

Table~\ref{tab:constants} summarizes the protocol constants used throughout AESP.

\begin{table}[h]
\centering
\caption{Protocol Constants}
\label{tab:constants}
\small
\begin{tabular}{@{}lll@{}}
\toprule
\textbf{Constant} & \textbf{Value} & \textbf{Section} \\
\midrule
REV32 context namespace & \texttt{ACEGF-REV32-V1-*} + context labels & \ref{sec:identity} \\
Max hierarchy depth & 5 & \ref{sec:identity} \\
Policy scope ranks & 1, 2, 3, 10 & \ref{sec:policy} \\
Max negotiation rounds & 10 & \ref{sec:negotiation} \\
Negotiation TTL & 24 hours & \ref{sec:negotiation} \\
Review deadline & 30 minutes & \ref{sec:review} \\
Address pool size & 5 & \ref{sec:privacy} \\
Consolidation interval & 4 hours & \ref{sec:privacy} \\
Consolidation jitter & ${\pm}$30\% & \ref{sec:privacy} \\
Consolidation batch & 5 & \ref{sec:privacy} \\
Inter-batch delay & 10--60 min & \ref{sec:privacy} \\
Audit batch threshold & 50 & \ref{sec:privacy} \\
Audit time window & 5 min & \ref{sec:privacy} \\
EIP-712 domain name & ``YalletAgentCommitment'' & \ref{sec:commitment} \\
EIP-712 version & ``1'' & \ref{sec:commitment} \\
Argon2id memory & 4 MB & \ref{sec:crypto} \\
Argon2id iterations & 3 & \ref{sec:crypto} \\
HKDF info prefix & \texttt{ACEGF-REV32-V1-} & \ref{sec:crypto} \\
Supported chains & 7 (see \S\ref{sec:rev32}) & \ref{sec:crypto} \\
\bottomrule
\end{tabular}
\end{table}

\section{Cryptographic Module API Summary}
\label{app:wasm_api}

Table~\ref{tab:wasm_api} lists the primary exports used by the AESP TypeScript SDK. Function names shown are from the WASM binding (\texttt{wasm.rs}); the \texttt{\_wasm} suffix is a binding-layer convention and does not indicate WASM-specific behavior---equivalent functions are exposed through the C~FFI (\texttt{ffi.rs}) and Dart~FFI bindings with their respective platform naming conventions (e.g., \texttt{\_ffi} suffix for C~FFI).

\begin{table}[h]
\centering
\caption{Primary Cryptographic Module Exports (WASM binding names)}
\label{tab:wasm_api}
\small
\begin{tabular}{@{}lp{4.5cm}@{}}
\toprule
\textbf{Export} & \textbf{Function} \\
\midrule
\texttt{view\_wallet\_unified\_with\_context\_wasm} & Derive multi-chain wallet under context \\
\texttt{evm\_sign\_typed\_data\_with\_context} & EIP-712 signing with context isolation \\
\texttt{solana\_sign\_transaction\_with\_context} & Solana transaction signing \\
\texttt{acegf\_encrypt\_for\_xidentity} & Encrypt data for a recipient identity \\
\texttt{acegf\_compute\_dh\_key\_wasm} & X25519 ECDH key agreement \\
\texttt{acegf\_seal\_passphrase\_wasm} & Argon2id passphrase sealing \\
\texttt{acegf\_unseal\_passphrase\_wasm} & Passphrase unsealing \\
\texttt{pqc\_sign\_message\_wasm} & ML-DSA-44 post-quantum signing \\
\texttt{pqc\_verify\_signature\_wasm} & ML-DSA-44 verification \\
\bottomrule
\end{tabular}
\end{table}

\end{document}